\documentclass[aps,prl,twocolumn,groupedaddress,amssymb,floatfix]{revtex4}
\usepackage{graphicx}

\newcommand{\sigvec}[0]{\mbox{\boldmath $\sigma$}}

\begin{document}


\title{Evidence for time-reversal symmetry breaking in the non-centrosymmetric superconductor LaNiC$_2$}


\author{A.D. Hillier, J. Quintanilla}
\affiliation{ISIS facility, STFC Rutherford Appleton Laboratory, Harwell Science and Innovation Campus, Oxfordshire, OX11 0QX, UK}
\author{R. Cywinski}
\affiliation{School of Applied Sciences, University of Huddersfield, Queensgate, Huddersfield, HD1 3DH, UK}

\date{\today}

\begin{abstract}
The results from muon spin relaxation experiments on the non-centrosymmetric intermetallic superconductor LaNiC$_2$ are reported. We find that the onset of superconductivity coincides with the appearance of spontaneous magnetic fields, implying that in the superconducting state time reversal symmetry is broken. An analysis of the possible pairing symmetries suggests only four triplet states compatible with this observation, all of which are non-unitary. They include the intriguing possibility of triplet pairing with the full point group symmetry of the crystal, which is only possible in a non-centrosymmetric superconductor. 
\end{abstract}


\maketitle

Symmetry breaking is a central concept of physics for which superconductivity provides a paradigm. In a conventional superconductor\cite{Bardeen57_2} gauge symmetry is broken, while unconventional superfluids and superconductors break other symmetries as well\cite{Sigrist91}. Examples include $^3$He\cite{Lee97}, cuprate high-temperature superconductors\cite{Annett90} and the ruthenate Sr$_2$RuO$_4$\cite{Mackenzie03}. Recently, superconductivity has been discovered in a number of materials whose lattices lack inversion symmetry\cite{Bauer04,Togano04,Akazawa04,Badica05,Tateiwa07}, with important implications for the symmetry of the superconducting state. However despite intense theoretical\cite{Gorkov01,Frigeri04,Sergienko04,Samokhin04,Hayashi06,Yanase07,Samokhin08} and experimental\cite{Bauer04,Togano04,Akazawa04,Badica05,Yuan06,Hafliger07,Tateiwa07,Fak08} efforts the issue of symmetry breaking in these systems remains uncertain. One of the most direct ways of detecting an unconventional superconducting state is muon spin relaxation ($\mu$SR), as it can unambiguously establish broken time reversal symmetry (TRS) \cite{Luke98,Aoki03}. In this letter we report $\mu$SR results on the non-centrosymmetric superconductor LaNiC$_2$ showing that TRS is broken on entering the superconducting state and we analyse  the possible symmetries. We identify four possibilities compatible with our observation, all of them non-unitary, including one where TRS is broken without concomitantly breaking any point symmetries of the crystal.
 
The RNiC$_2$ (where R= rare earth) intermetallic alloys were first reported by Bodak and Marusin\cite{Bodak79}. Interestingly they crystallise with the non-centrosymmetric space group Amm2. For different R these alloys exhibit a wide range of magnetic ground states. However, LaNiC$_2$ does not order magnetically but exhibits superconductivity at 2.7~K\cite{Lee96,Pecharsky98}. Deviations from conventional BCS behaviour have been interpreted as evidence for triplet superconductivity\cite{Lee96}. 

Evidence for unconventional pairing can be shown from time-reversal symmetry (TRS) breaking through the detection of spontaneous but very small internal fields\cite{Sigrist91}. $\mu$SR is especially sensitive for detecting small changes in internal fields and can easily measure fields of 0.1~G which corresponds to $\approx$0.01~$\mu_B$. This makes $\mu$SR an extremely powerful technique for measuring the effects of TRS breaking in exotic superconductors. Direct observation of TRS breaking states is extremely rare and spontaneous fields have only been observed in a few systems: PrOs$_4$Sb$_{12}$\cite{Aoki03}, Sr$_2$RuO$_4$\cite{Luke98}, B-phase of UPt$_3$\cite{Luke93} and (U,Th)Be$_{13}$\cite{Heffner90}. 

The sample was prepared by melting together stoichiometric amounts of the constituent elements in a water-cooled argon arc furnace. Part of the sample was crushed into a fine powder and characterised by neutron powder diffraction using D1B at the Institute Laue Langevin, Grenoble, France. The $\mu$SR experiments were carried out using the MuSR spectrometer in longitudinal geometry. At the ISIS facility, a pulse of muons is produced every 20 ms and has a FWHM of $\sim$70 ns. These muons are implanted into the sample and decay with a half-life of 2.2~$\mu$s into a positron which is emitted preferentially in the direction of the muon spin axis. These positrons are detected and time stamped in the detectors which are positioned either before, F, and after, B, the sample for longitudinal (relaxation) experiments. Using these counts the asymmetry in the position emission can be determined and, therefore, the muon polarisation is measured as a function of time. The sample was a thin disk, 30~mm in diameter and 1 mm thick, mounted onto a 99.995+~$\%$ pure silver plate. Any muons stopped in silver give a time independent background for longitudinal (relaxation) experiments. The sample holder and sample were mounted onto a TBT dilution fridge with a temperature range of 0.045-4~K. The sample was cooled to base temperature in zero field and the $\mu$SR spectra were collected upon warming the sample while still in zero field. The stray fields at the sample position are cancelled to within 1~$\mu$T by a flux-gate magnetometer and an active compensation system controlling the three pairs of correction coils, then the sample was cooled to base temperature in a longitudinal field of 5~mT and the $\mu$SR spectra were collected while warming.

\begin{figure}
\includegraphics[width=8cm,height=6cm]{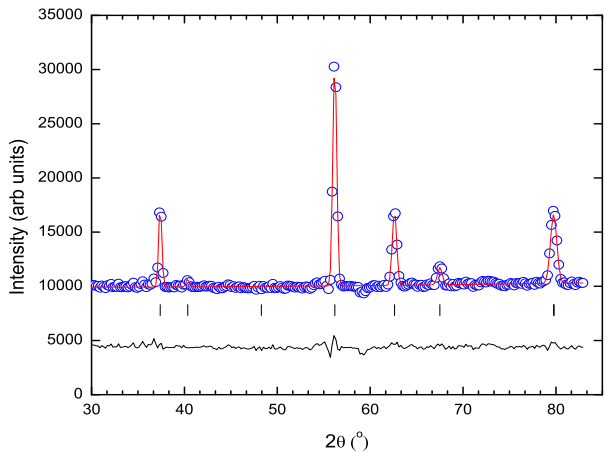}%
\caption{\label{fig:diffpatt} (color online) The powder neutron diffraction pattern of LaNiC$_2$ at 300~K. The line is a Rietveld refinement to the data. The vertical tickmarks indicate the calculated peak positions and the lower graph shows the difference plot.}
\end{figure}

The powder neutron diffraction measurements confirmed the sample had crystallised into a single phase of the expected orthorhombic space group Amm2 with lattice parameters of a=3.96~\r{A} and b=4.58~\r{A} and c=6.20~\r{A} (see Fig.~\ref{fig:diffpatt}). These results agree well with the current literature. The point group mm2 ($C_{2v}$) has a particularly low symmetry with only 4 irreducible representations, all of them one-dimensional. The characters of the 4 symmetry operations in each of the 4 (one-dimensional) irreducible representations are given in Table \ref{tab:C2v}. We also give two alternative basis functions for each irreducible represenation: one even under inversion and one odd (this would not be possible for a centrosymmetric system, where even and odd basis functions belong to distinct irreducible representations). We use the notation of Ref.~\cite{Annett90}, whereby $X$ represents any function of ${\bf k}$ that is continuous through the Brillouin zone boundary and transforms like $k_x$ under the symmetry operations in the point group.

\begin{table}
\begin{tabular}{|c|c|c|c|c|c|c|}
\hline 
\textbf{$C_{2v}$}&
\multicolumn{4}{c|}{\begin{tabular}{c}
symmetries and\tabularnewline
their characters\tabularnewline
\end{tabular}}&
\multicolumn{2}{c|}{\begin{tabular}{c}
sample\tabularnewline
basis functions\tabularnewline
\end{tabular}}\tabularnewline
\hline 
\begin{tabular}{c}
irreducible\tabularnewline
representations\tabularnewline
\end{tabular}&
$E$&
$C_{2}$&
$\sigma_{v}$&
$\sigma_{v}'$&
even&
odd\tabularnewline
\hline 
$A_{1}$&
1&
1&
1&
1&
$1$&
$Z$\tabularnewline
\hline 
$A_{2}$&
1&
1&
-1&
-1&
$X Y$&
$X Y Z$\tabularnewline
\hline 
$B_{1}$&
1&
-1&
1&
-1&
$X Z$&
$X$\tabularnewline
\hline 
$B_{2}$&
1&
-1&
-1&
1&
$Y Z$&
$Y$\tabularnewline
\hline
\end{tabular}
\caption{\label{tab:C2v} The character table of the point group of the crystal, $mm2$ ($C_{2v}$). The last two columns give two simple basis functions for each irreducible representation: one even (compatible with singlet pairing), and one odd (for triplet pairing). }
\end{table}

\begin{figure}
\includegraphics[width=7cm,height=5.5cm]{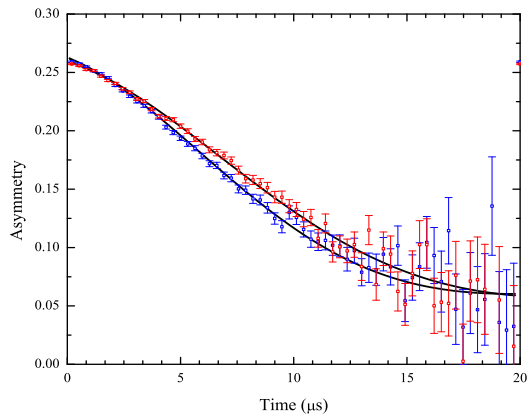}%
\caption{\label{fig:spectra} (color online) The zero field $\mu$SR spectra for LaNiC$_2$.  The blue symbols are the data collected at 54~mK and the red symbols are the data collected at 3.0~K. The lines are a least squares fit to the data.}
\end{figure}

The absence of a  precessional signal in the $\mu$SR spectra at all temperatures confirms that there are no spontaneous coherent internal magnetic fields associated with long range magnetic order in LaNiC$_2$ at any temperature. In the absence of atomic moments muon spin relaxation is expected to arise entirely from the local fields associated with the nuclear moments. These nuclear spins are static, on the time scale of the muon precession, and randomly orientated. The depolarisation function, $G_z(t)$, can be described by the Kubo-Toyabe function\cite{Hayano79}
\begin{equation}
\label{eq:fitfun}
G_z^{KT}(t)=(\frac{1}{3}+\frac{2}{3}(1-\sigma ^2 t^2)\exp(-\frac{\sigma^2t^2}{2})),
\end{equation}

\noindent where $\sigma/\gamma_\mu$ is the local field distribution width and  $\gamma_\mu$=13.55MHz/T is the muon gyromagnetic ratio. The spectra that we find for LaNiC$_2$ are well described by the function
\begin{equation}
\label{eq:fitfun2}
G_z(t)=A_0G_z^{KT}(t)\exp(-\lambda t) + \\
A_{bckgrd},
\end{equation}

\noindent where $A_0$ is the initial asymmetry, A$_{bckgrd}$ is the background, and $\lambda$ is the electronic relaxation rate (see Fig. \ref{fig:spectra}). 
It is assumed that the exponential factor involving $\lambda$ arises from electronic moments which afford an entirely independent muon spin relaxation channel in real time. This term will be discussed in greater detail later.

\begin{figure}
\includegraphics[width=5cm,angle=-90]{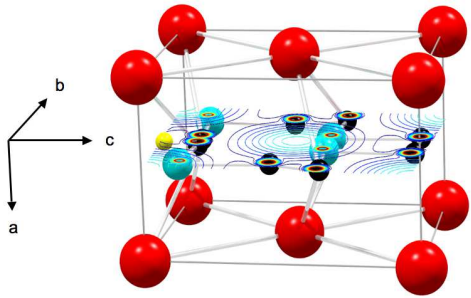}
\caption{\label{fig:calc} (color online) The crystal structure of LaNiC$_2$. The large red spheres are La, medium size blue spheres are Ni, smaller black spheres are C and the smallest yellow sphere is the muon. The contour plot of the nuclear dipole fields for LaNiC$_2$ is shown throughout the unit cell at x=$\frac{1}{2}$ and shows the muon site is at $\left(\frac{1}{2},\frac{1}{2},0\right)$. }
\end{figure}

The coefficients A$_0$, $\sigma$, and A$_{bckgrd}$ are found to be temperature independent. The contribution from the nuclear moments was found to be $\sigma$=0.08~$\mu$s$^{-1}$. Finite element analysis\cite{Hillier07} indicates that the observed sigma is consistent with the muon localising at the 1/2,1/2,0 and equivalent sites (see Figure~\ref{fig:calc}). However we emphasize that our conclusions do not rely on the muon being on any particular location.  

\begin{figure}
\includegraphics[width=7cm]{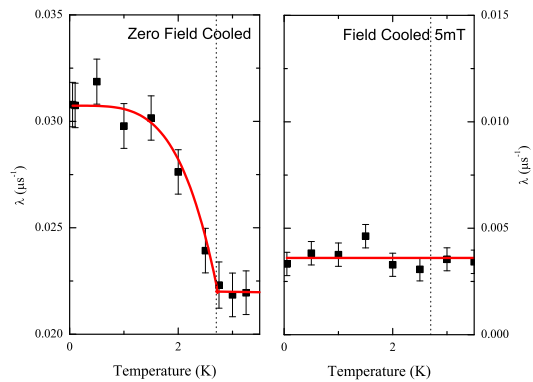}
\caption{\label{fig:LambdavT} (color online) The left figure is the temperature dependence of the electronic relaxation rate, $\lambda$, for LaNiC$_2$ in zero-field, which clearly shows the spontaneous fields appearing at T$_C$=2.7~K \cite{Lee96,Pecharsky98}(dotted line). The right figure is the temperature dependence of $\lambda$ for an applied field of 5~mT, in which a flat temperature dependence is observed. The red lines are guides to the eye.} 
\end{figure}

The only parameter that shows any temperature dependence is $\lambda$, which increases rapidly with decreasing temperature below T$_C$(see Fig. \ref{fig:LambdavT}~a). As indicated earlier, it is most probable that the exponential relaxation process of equation \ref{eq:fitfun2} arises from the field distribution associated with electronic rather than nuclear spins. Indeed, such an exponential form is generally attributed to magnetic fields of atomic origin that are fluctuating sufficiently rapidly for their effects on the muon relaxation to be motionally narrowed. However in this case we find that a weak longitudinal magnetic field of only 5~mT is sufficient to fully decouple the muon from this relaxation channel(see Fig. \ref{fig:LambdavT}~b). This in turn suggests that the associated magnetic fields are in fact static or quasistatic on the time scale of the muon precession. In a superconductor with broken TRS, spontaneous magnetic fields arise in regions where the order parameter is inhomogeneous, such as domain walls and grain boundaries\cite{Sigrist91}. Thus, the appearance of such spontaneous static fields at T$_C$ provides convincing evidence for time reversal symmetry breaking on entering the superconducting state of LaNiC$_2$.

We have been able preclude the possibility of the observed effects arising from magnetic impurities within the sample. Although, in principle, very small amounts of parasitic rare earth impurity ions ($<0.05\%$) associated with 
the La metal may be present, the dynamic dipolar fields arising from such dilute impurities would contribute either a motionally narrowed lorentzian Kubo Toyabe, or a root exponential term to the muon relaxation. In any event the associated relaxation rate would be expected to vary smoothly with temperature. Any residual ferromagnetic impurity phases (e.g. Ni) would contribute only to a very small ($\ll 1\%$) reduction of the initial asymmetry. Finally, a small amount of LaC$_2$ impurity would not give any signal at 2.7~K as LaC$_2$ is a non-magnetic superconductor ($T_C$=1.6~K) \cite{Sakai81}.

Let us now discuss the implications of this result for the pairing symmetry. We address in detail just the simplest case where the superconducting state does not break translational symmetry and moreover spin-orbit coupling (SOC) does not play a role. Just below the superconducting instability, very general group-theoretical arguments \cite{Annett90} require that the gap matrix $\hat{\Delta}({\bf k})$ correspond to one of the irreducible representations of the space group of the crystal. In the simplest case, the group to be considered is the direct product $SO(3) \times C_{2v}$. The group $SO(3)$ corresponds to arbitrary rotations in spin space, and has two irreducible representations corresponding to singlet pairing (of dimension 1) and triplet pairing (dimension 3);  $C_{2v}$ is the point group of the crystal and it has the four one-dimensional representations given in Table \ref{tab:C2v}. This gives a total of 8 irreducible representations of the product group: 4 one-dimensional, singlet representations and 4 three-dimensional, triplet representations. The latter have an order parameter that transforms like a vector under spin rotations. The two possible ground states in a general Ginzburg-Landau theory of that symmetry are given in Ref.~\cite{Annett90}. This leads to the 12 possible order parameters given in Table \ref{tab:states}, where we have used the freedom to choose a basis function of either even or odd symmetry for each irreducible representation of the point group (available only in non-centrosymmetric crystals) to satisfy Pauli's exclusion principle $\Delta_{\sigma,\sigma'}({\bf k}) = -\Delta_{\sigma',\sigma}(-{\bf k})$ in all cases. Of these 12, 8 are unitary states (4 spin singlets and 4 spin triplets) while the other 4 are non-unitary. Only those 4 have non-trivially complex order parameters and thus break TRS. In them, only spin-up electrons participate in pairing, and therefore there is an ungapped Fermi surface coexisting with another that has one ($^3A_1,^3B_1,^3B_2$) or three ($^3A_2$) line nodes. This would suggest a specific heat $C \sim \gamma_{\downarrow} T + \gamma_{\uparrow} T^2$ at low temperatures \cite{Sigrist91}, which is at odds with $C \sim T^3$ reported in Ref.~\cite{Lee96} (compatible with both Fermi surfaces having point nodes \cite{Sigrist91}). Nevertheless, we note that in the case of another triplet superconductor, Sr$_2$RuO$_4$, the low-temperature thermodynamics remained unclear until large, high-purity single crystals were available \cite{Mackenzie03}. 

The fact that all TRS breaking states are non-unitary is a direct consequence of the particularly low crystal symmetry (c.f. D$_{4h}$, the point group relevant to Sr$_2$RuO$_4$, where there are both unitary and non-unitary order parameters that break TRS - indeed unitary pairing is realised in that system \cite{Mackenzie03}). Of these four, the one corresponding to $^3A_1$ breaks gauge and time reversal symmetries only, unlike those corresponding to $^3A_2$, $^3B_1$ and $^3B_2$ which break additional symmetries of the crystal. The possibility of breaking gauge and time reversal symmetries without concomitantly breaking other symmetries is unique to non-centrosymmetric superconductors. (This is consistent with the observation that the superconducting order parameter of a non-centrosymmetric supercondcutor can always have a triplet component, even when only gauge symmetry is broken \cite{Yuan06}.) Again, we expect that establishing this possibility will require the availability of large single crystals. 

The above analysis neglected the effect of strong SOC on the superconducting state. If SOC is strong, the point group to consider is obtained by appending to each rotation element a simultaneous rotation of the spin \cite{Annett90}. This is especially relevant in the case of non-centrosymmetric systems as the SOC terms in the Hamiltonian mix the singlet and triplet components of the order parameter and can alter the superconducting state dramatically
 \cite{Gorkov01,Frigeri04,Sergienko04,Samokhin08}. In particular, a Rashba SOC term tends to suppress triplet pairing \cite{Frigeri04}. Which triplet components survive is dictated by details of the pairing interaction and band structure \cite{Samokhin08}. SOC is believed to be crucial in CePt$_3$Si and Li$_2$Pt$_3$B, but not in Li$_2$Pd$_3$B \cite{Yuan06}. Further work will be required to ascertain the possible importance of SOC in LaNiC$_2$. 

\begin{table}
  \begin{tabular}{|c|c|c|}
    \hline
    \textbf{$SO(3) \times C_{2v}$}
    & 
    \begin{tabular}{c}
      gap function
      \tabularnewline
      (unitary)
    \end{tabular}
    &
    \begin{tabular}{c}
      gap function
      \tabularnewline
      (non-unitary)
    \end{tabular}
    \tabularnewline
    \hline
    \hline
    $^1A_{1}$
    &
    $\Delta({\bf k}) = 1$
    &
    -
    \tabularnewline
    $^1A_{2}$
    &
    $\Delta({\bf k}) = XY$
    &
    -
    \tabularnewline
    $^1B_{1}$
    &
    $\Delta({\bf k}) = XZ$
    &
    -
    \tabularnewline
    $^1B_{2}$
    &
    $\Delta({\bf k}) = YZ$
    &
    -
    \tabularnewline
    \hline  
    $^3A_{1}$
    &
    ${\bf d}({\bf k}) = (0,0,1)Z$
    &
    ${\bf d}({\bf k}) = (1,i,0)Z$
    \tabularnewline
    $^3A_{2}$
    &
    ${\bf d}({\bf k}) = (0,0,1)XYZ$
    &
    ${\bf d}({\bf k}) = (1,i,0)XYZ$
    \tabularnewline
    $^3B_{1}$
    &
    ${\bf d}({\bf k}) = (0,0,1)X$
    &
    ${\bf d}({\bf k}) = (1,i,0)X$
    \tabularnewline
    $^3B_{2}$
    &
    ${\bf d}({\bf k}) = (0,0,1)Y$
    &
    ${\bf d}({\bf k}) = (1,i,0)Y$
    \tabularnewline
    \hline
  \end{tabular}
  \caption{\label{tab:states}Homogeneous superconducting states allowed by symmetry, for weak spin-orbit coupling. We have used the standard notation $\hat{\Delta}({\bf k})=\Delta({\bf k})i\hat{\sigma}_y$ for singlet states and $\hat{\Delta}({\bf k})=i\left[{\bf d}({\bf k}).\hat{\sigvec}\right]\hat{\sigma}_y$ for triplets, where $\hat{\sigvec}\equiv (\hat{\sigma}_x,\hat{\sigma}_y,\hat{\sigma}_z)$ is the vector of Pauli matrices. Only the four non-unitary triplet states are compatible with our observation of broken TRS.}
\end{table}

In conclusion, zero field and longitudinal field $\mu$SR experiments have been carried out on LaNiC$_2$. The zero field measurements show a spontaneous field appearing at the superconducting transition temperature (T$_C$=2.7~K\cite{Lee96,Pecharsky98}). The application of a 5~mT longitudinal field shows that these spontaneous fields are static or quasi-static on the time scale of the muon. This provides convincing evidence that time reversal symmetry is broken in the superconducting state of LaNiC$_2$. Thus this material is the first non-centrosymmetric superconductor to join the ranks of Sr$_2$RuO$_4$, PrOs$_4$Sb$_{12}$, the B-phase of UPt$_3$, (U,Th)Be$_{13}$ and also the A or A1 phase of superfluid $^3$He. Compared to these systems, non-centrosymmetric superconductors are expected to have many novel properties, such as field-tuned Fermi surface topology\cite{Eremin06} and topologically protected spin currents\cite{Tanaka08}. Our analysis of the possible pairing symmetries in LaNiC$_2$ suggests four possible triplet states, all of them non-unitary (analogous to the A1 phase of $^3$He\cite{Lee97}) featuring one unpaired Fermi surface and another one with line nodes, and including the intriguing possibility of triplet pairing with the full point group symmetry of the crystal, which is possible only for non-centrosymmetric superconductors.

\begin{acknowledgments}
The authors would like to acknowledge useful discussions with J.F. Annett, S.T. Carr, N. Gidopoulos, B.L. Gy\"orffy and B.J. Powell. J. Quintanilla acknowledges support by CCLRC (now STFC) in association with St. Catherine's College, Oxford.
\end{acknowledgments}


\end{document}